# Mass Generation in Structural Algebraic Quantum Field Theory: An alternative to the Higgs mechanism

A. D. Alhaidari

*Saudi Center for Theoretical Physics, P.O. Box 32741, Jeddah 21438, Saudi Arabia*

**Abstract:** Within the recently proposed structure-inclusive algebraic formulation of quantum field theory, we show that a massless particle can acquire mass by special nonlinear coupling to a universal massless scalar field; establishing an alternative to the Higgs mechanism in the standard model of particle physics. We end with a conjecture concerning dark energy and dark matter.

## 1. Introduction

Conventional quantum field theory (QFT) describes structureless elementary particles and their interaction [1-3]. However, some persistent difficulties in the theory motivated the development of few alternatives and several attempts at modifications (see, for example, Ref. [4,5]). These difficulties prompted various creative solutions that included renormalization, confinement, topological methods, asymptotic freedom, algebraic methods, etc. One of our main concerns is the inadequacy of conventional QFT to describe particles with infinitesimal structure at very low energies (e.g., nucleons at KeV energy) because at this energy scale such a structure should have no significant contribution and particles are virtually structureless; hence conventional QFT should produce reasonably accurate results, but it doesn't. Recently, we proposed an algebraic formulation of quantum field theory that incorporates particles with structure [6,7]. We refer to it as "Structural Algebraic Quantum Field Theory" with the acronym/ initialism SAQFT. The theory of special functions and orthogonal polynomials [8-11] play central role in the development of the formulation, which has two main points of departure from conventional QFT. The first, is that the free quantum fields are resolved by means of a Fourier expansion in the energy rather than linear momentum. The second point, is that particle structure is built into the formulation. Justification for, and validation of, these two points were demonstrated therein [6,7]. However, if it turns out that the particle is indeed structureless for all energies then that could easily be accommodated by setting the structure to zero making SAQFT equivalent to conventional QFT. A remarkable property of the theory is that we can construct physical models in which closed loops in the Feynman diagrams are finite doing away with renormalization. One such model was, in fact, presented in Sec. 4 of Ref. [6,7].

In this article, we show how a massless particle in SAQFT can acquire mass by a particular nonlinear coupling to a special massless scalar particle giving an alternative to the Higgs mechanism in conventional QFT of elementary particle physics. A brief, simple, and nontechnical introduction to the Higgs mechanism can be found in some popular books like that of Ref. [12]. On the other hand, specialized technical details can be found in many books on quantum field theory (See, for example, Ref. [13] and citations therein). Nevertheless, several alternatives to the Higgs mechanism have already been suggested (see, for example, [14-19] and references cited therein). We start by giving a brief account of the formulation of SAQFT using scalar particles as an illustration and as benefit to the reader.

In the relativistic units $\hbar = c = 1$, the quantum field associated with a scalar particle in 3+1 Minkowski space-time is written in SAQFT as follows

$$\Psi(t,\vec{r}) = \int_{\Omega} e^{-iEt}\psi(E,\vec{r})\,\boldsymbol{a}(E)\,dE + \sum_{j=0}^{N} e^{-iE_j t}\psi_j(\vec{r})\,\boldsymbol{a}_j \,, \tag{1}$$

where $E^2 \geq M^2$ for $E \in \Omega$ and $0 \leq E_j^2 < M^2$ with $M$ being the rest mass of the particle. Therefore, for massless particles, the sum in (1) (i.e., the structure) is absent. This implies that all massless particles are structureless. However, the reverse is not true. That is, not all structureless particles are massless as it is evidently clear in conventional QFT (e.g., the electron). However, all particles with internal structure [i.e., non-vanishing sum in (1)] are massive. The creation/annihilation operators satisfy the conventional commutation algebra,

$$\left[\boldsymbol{a}(E),\boldsymbol{a}^{\dagger}(E')\right] := \boldsymbol{a}(E)\boldsymbol{a}^{\dagger}(E') - \boldsymbol{a}^{\dagger}(E')\boldsymbol{a}(E) = \delta(E-E')\,, \qquad \left[\boldsymbol{a}_i,\boldsymbol{a}_j^{\dagger}\right] = \lambda^{-1}\delta_{i,j}\,. \tag{2}$$

where $\lambda$ is a real positive scale parameter of inverse length dimension that gives a measure of the structure size and/or mass. All other commutators among $\boldsymbol{a}(E)$, $\boldsymbol{a}^{\dagger}(E)$, $\boldsymbol{a}_j$, and $\boldsymbol{a}_j^{\dagger}$ vanish. The continuous and discrete Fourier kernels are written as the following pointwise convergent series

$$\psi(E,\vec{r}) = \sum_{n=0}^{\infty} f_n(E)\phi_n(\vec{r}) = f_0(E)\sum_{n=0}^{\infty} p_n(z)\phi_n(\vec{r})\,, \tag{3a}$$

$$\psi_j(\vec{r}) = \sum_{n=0}^{\infty} g_n(E_j)\phi_n(\vec{r}) = g_0(E_j)\sum_{n=0}^{\infty} p_n(z_j)\phi_n(\vec{r})\,. \tag{3b}$$

where $z$ is an energy-dependent parameter (the *spectral parameter*) and $\{f_n, g_n\}$ are real expansion coefficients which are written as $f_n = f_0 p_n$ and $g_n = g_0 p_n$ making $p_0 = 1$. $\{\phi_n(\vec{r})\}$ is a complete set of functions in configuration space that belong to the solution space of the free wave equation. $\{p_n(z)\}$ is a sequence of orthogonal polynomials in $z$ (the *spectral polynomials*) that satisfy the following symmetric three-term recursion relation [6,7]

$$z\,p_n(z) = \alpha_n p_n(z) + \beta_{n-1}p_{n-1}(z) + \beta_n p_{n+1}(z)\,, \tag{4}$$

for $n = 1,2,3,...$ and with the two initial values $p_0(z) = 1$ and $p_1(z) = (z-\alpha_0)/\beta_0$ *. The recursion coefficients $\{\alpha_n, \beta_n\}$ are real constants that are independent of $z$ and such that $\beta_n \neq 0$ for all $n$. Additionally, the spectral polynomials satisfy the following generalized orthogonality [6-11]

$$\int_{\Omega} \rho(z)p_n(z)p_m(z)\,dz + \sum_{j=0}^{N} \varsigma_j\, p_n(z_j)p_m(z_j) = \delta_{n,m}\,, \tag{5}$$

* Generally, $p_1(z)$ is a two-parameter linear function of $z$.

where $\rho(z)$ is the continuous component of the weight function and $\varsigma_j$ is the discrete component. Using the recursion relation (4) with $z=(E^2-M^2)/\lambda^2$, the free Klein-Gordon wave equation $\left(\partial_t^2-\vec{\nabla}^2+M^2\right)\Psi(t,\vec{r})=0$ gives

$$-\vec{\nabla}^2\phi_n(\vec{r})=\lambda^2\left[\alpha_n\phi_n(\vec{r})+\beta_{n-1}\phi_{n-1}(\vec{r})+\beta_n\phi_{n+1}(\vec{r})\right], \tag{6}$$

Therefore, the *algebraic* three-term recursion relation (4) is equivalent to the *differential* free wave equation. The quantum field conjugate to $\Psi(t,\vec{r})$ is written as

$$\bar{\Psi}(t,\vec{r})=\int_\Omega e^{iEt}\bar{\psi}(E,\vec{r})\boldsymbol{a}^\dagger(E)\,dE+\sum_{j=0}^{N}e^{iE_jt}\bar{\psi}_j(\vec{r})\boldsymbol{a}_j^\dagger, \tag{7}$$

where the conjugate Fourier kernels $\bar{\psi}(E,\vec{r})$ and $\bar{\psi}_j(\vec{r})$ are identical to (3) but with $\phi_n(\vec{r})\mapsto\bar{\phi}_n(\vec{r})$ such that

$$\left\langle\phi_n(\vec{r})\middle|\bar{\phi}_m(\vec{r})\right\rangle=\left\langle\bar{\phi}_n(\vec{r})\middle|\phi_m(\vec{r})\right\rangle=\frac{1}{\lambda}\delta_{n,m}, \tag{8a}$$

$$\sum_{n=0}^{\infty}\phi_n(\vec{r})\bar{\phi}_n(\vec{r}')=\sum_{n=0}^{\infty}\bar{\phi}_n(\vec{r})\phi_n(\vec{r}')=\frac{1}{\lambda}\delta^3(\vec{r}-\vec{r}'), \tag{8b}$$

Consequently, $i\lambda\bar{\Psi}(t,\vec{r})$ is the canonical conjugate to $\Psi(t,\vec{r})$ because if we take $f_0^2(E)dE=\rho(z)dz$ and $g_0^2(E_j)=\lambda\varsigma_j$ then their equal-time commutation relation becomes [6,7]

$$\left[\Psi(t,\vec{r}),i\lambda\bar{\Psi}(t,\vec{r}')\right]=i\delta^3(\vec{r}-\vec{r}'), \tag{9}$$

where we have used the commutation algebra (2), orthogonality (5), and completeness (8b).

A neutral scalar particle is represented by the quantum field $\Phi(t,\vec{r})=\frac{1}{\sqrt{2}}\left[\Psi(t,\vec{r})+\bar{\Psi}(t,\vec{r})\right]$ with $\bar{\phi}_n(\vec{r})=\phi_n^\dagger(\vec{r})$. On the other hand, the complex (charged) scalar particle is defined by the positive-energy quantum field

$$\Phi(t,\vec{r})=\frac{1}{\sqrt{2}}\left[\Psi_+(t,\vec{r})+\Psi_-^\dagger(t,\vec{r})\right], \tag{10a}$$

where $\Psi_\pm(t,\vec{r})$ is identical to (1) but with the associated spectral polynomials $\left\{p_n^\pm(z)\right\}$ along with their recursion coefficients $\left\{\alpha_n^\pm,\beta_n^\pm\right\}$, and with annihilation operators $\boldsymbol{a}_\pm(E)$ and $\boldsymbol{a}_\pm^j$ such that $\left[\boldsymbol{a}_r(E),\boldsymbol{a}_{r'}^\dagger(E')\right]=\delta_{r,r'}\,\delta(E-E')$ and $\left[\boldsymbol{a}_r^i,(\boldsymbol{a}_{r'}^j)^\dagger\right]=\lambda^{-1}\delta_{r,r'}\,\delta^{i,j}$ where $r$ and $r'$ stand for ±. The corresponding charged scalar antiparticle is represented by the following negative-energy quantum field

$$\bar{\Phi}(t,\vec{r})=\frac{1}{\sqrt{2}}\left[\bar{\Psi}_+(t,\vec{r})+\bar{\Psi}_-^\dagger(t,\vec{r})\right]. \tag{10b}$$

Spinors in SAQFT could also be constructed as shown in Appendix B of Ref. [6,7]. The Feynman propagators for these quantum fields are obtained as described in [6,7].

In Section 2, we introduce the proposed scheme for generating mass by a special nonlinear coupling to a massless scalar particle, which we refer to as the "universal particle/field". It is the analog of the Higgs particle but with the major difference that the universal particle is massless whereas the Higgs is very massive (about 125 GeV). Our conclusion is given in Section 3 where we end by putting forth a *conjecture* regarding dark energy and dark matter.

## 2. The proposed scheme

It should be evident from the above analysis and from the formulation of SAQFT introduced in [6,7] that all physical properties of a particle are obtained from those of the associated spectral polynomials $\left\{p_n(z)\right\}$. Therefore, an interaction in SAQFT (for example, inelastic scattering) can change these spectral polynomials giving a new recursion relation and orthogonality that replace (4) and (5), respectively. The wave equation that embodies the interaction results in a modified recursion relation for these polynomials that reads

$$z\, p_n(z) = \alpha_n p_n(z) + \beta_{n-1} p_{n-1}(z) + \beta_n p_{n+1}(z) + \Delta p_n(z)\,, \tag{11}$$

where $\Delta p_n(z)$ is due to the interaction added to the free wave equation. Here, we propose a nonlinear interaction Lagrangian model that reads $\mathscr{L}_I = \eta\otimes\Phi(\bar{\Theta}\Theta) + \frac{1}{2}\kappa\otimes(\bar{\Phi}\Phi)^2$, where $\Phi$ is the universal scalar field, $\Theta$ is the spinor field, and $\{\eta,\kappa\}$ are dimensionless coupling tensors. The positive energy component of this interaction Lagrangian model reads as follows (where for simplicity of the presentation, we show neutral particles and include a single spin projection):

$$\eta\otimes\Phi(\bar{\Theta}\Theta) = \sum_{n,a,b=0}^{\infty}\eta_{n,a,b}\left[\int_\Omega e^{-i(E-E'+E'')t}dz\,dz'dz''\sqrt{\rho(z)\omega(z')\omega(z'')}\; p_n(z)q_a(z')q_b(z'')\phi_n(\vec{r})\bar{\vartheta}_a(\vec{r})\vartheta_b(\vec{r})\boldsymbol{a}(E)\boldsymbol{b}^\dagger(E')\boldsymbol{b}(E'')\right] \tag{12a}$$

$$\kappa\otimes(\bar{\Phi}\Phi)^2 = \sum_{n,m,k,l=0}^{\infty}\kappa_{n,m,k,l}\left[\int_\Omega e^{-i(E+E'-E''-E''')t}dz\,dz'dz''dz'''\sqrt{\rho(z)\rho(z')\rho(z'')\rho(z''')}\; p_n(z)p_m(z')p_k(z'')p_l(z''')\phi_n(\vec{r})\phi_m(\vec{r})\bar{\phi}_k(\vec{r})\bar{\phi}_l(\vec{r})\boldsymbol{a}(E)\boldsymbol{a}(E')\boldsymbol{a}^\dagger(E'')\boldsymbol{a}^\dagger(E''')\right] \tag{12b}$$

$\left\{\eta_{n,a,b}\right\}$ and $\left\{\kappa_{i,j,k,l}\right\}$ are elements of the coupling tensors. The spectral polynomials $\left\{p_n(z)\right\}_{n=0}^{\infty}$ are associated with the universal scalar $\Phi$ whereas $\left\{q_a(z)\right\}_{a=0}^{\infty}$ are the spectral polynomials associated with the spinor $\Theta$. Moreover, $\omega(z)$ is the continuous weight function associated with $\left\{q_a(z)\right\}$ and $\vartheta_a = \begin{pmatrix}\vartheta_a^+\\ \vartheta_a^-\end{pmatrix}$ is the corresponding two-component spinor basis functions that satisfy equations analogous to (6) and (8) for the spinor field [see Appendix B of Ref. [6,7] for details on the construction of spinor quantum fields in SAQFT; especially equations (B.9) and (B.17)]. The indices $\{a,b,c,..\}$ stand for the degrees of the spectral polynomials associated with the spinor field $\Theta$ whereas the indices $\{i,j,k,..\}$ refer to the polynomial degrees for the universal scalar field $\Phi$. Since the *free* quantum fields $\Phi$ and $\Theta$ represent *massless* particles then their respective spectral polynomials $\{p_n(z)\}$ and $\{q_a(z)\}$ satisfying the recursion relation (4) with their own recursion coefficients and with $z = E^2/\lambda^2$. Hence, these polynomials must have pure continuous spectra. That is, the sum in the quantum field (1) (i.e., the structure) and the sum in the orthogonality (5) are absent.

In the presence of the nonlinear interaction $\mathscr{L}_I$, the wave equation for the spinor quantum field $\Theta$ contains the term $\eta\otimes\Phi\Theta$ added to the massless free Dirac equation giving rise to a new orthogonal polynomial $q'_a(z)$ that differs from $q_a(z)$ in its recursion relation by the term $\Delta q_a(z)$. In the space of spectral polynomials and with a representation of the term $\eta\otimes\Phi\Theta$ that parallels (12a), this change is written as weighted sum of products of the polynomials $p_n(z)$ and $q_a(z)$. That is, in Eq. (11) for $q_b(z)$ we write

$$\Delta q_b(z) = \sum_{n,a=0}^{\infty} \eta_{n,a,b}\, p_n(z) q_a(z). \tag{13}$$

We can use the well-known linearization technique in orthogonal polynomials [8-11,20] to write

$$p_n(z) q_a(z) = \sum_{c=0}^{n+a} \boldsymbol{e}_{n,a,c}\, q_c(z), \tag{14}$$

where $\{\boldsymbol{e}_{n,a,c}\}$ are constants called the "linearization coefficients" (see Chap. 9 in the book [11] and Ref. [20]). Specifically, using results in Sec. 6 of Ref. [20], we can write

$$\boldsymbol{e}_{n,a,c} = \left[p_n(K)\right]_{a,c} = \boldsymbol{e}_{n,c,a}, \tag{15}$$

where $K$ is the tridiagonal symmetric matrix, whose elements are $K_{a,b} = \chi_a \delta_{a,b} + \xi_{a-1}\delta_{a-1,b} + \xi_a \delta_{a+1,b}$ with $\{\chi_a, \xi_a\}$ being the recursion coefficients associated with the spectral polynomials $\{q_a(z)\}$. Thus, Eq. (13) becomes $\Delta q_b(z) = \sum_{c=0}^{\infty} \Pi_{b,c} q_c(z)$, where $\Pi_{b,c} = \sum_{n,a=0}^{\infty} \eta_{n,a,b} \boldsymbol{e}_{n,a,c}$. For a given pair of massless particles $\Phi$ and $\Theta$ (i.e., a given pair of spectral polynomials $\{p_n(z)\}$ and $\{q_a(z)\}$ with continuous spectra) we choose the coupling elements of the vertex tensor, $\{\eta_{n,a,b}\}$, such that $\Pi$ becomes a tridiagonal symmetric matrix with a constant positive definite offset on the diagonal. That is, $\Pi_{a,b} = \left(\tilde{\chi}_a + \zeta^2\right)\delta_{a,b} + \tilde{\xi}_{a-1}\delta_{a-1,b} + \tilde{\xi}_a \delta_{a+1,b}$. Therefore, Eq. (11) for $q_a(z)$ and Eq. (13) show that the modified spectral polynomial $\{q'_a(z)\}$ associated with the spinor particle $\Theta$ satisfies a three-term recursion relation whose spectral parameter is $z = (E^2/\lambda^2) - \zeta^2$ and with recursion coefficients $\{\chi'_a, \xi'_a\}$ where $\chi'_a = \chi_a + \tilde{\chi}_a$ and $\xi'_a = \xi_a + \tilde{\xi}_a$. Therefore, the term $\lambda\zeta$ becomes the acquired mass for the spinor particle $\Theta$. In practice, however, the desired massive spinor particle is physically pre-determined giving the associated spectral polynomial $q'_a(z)$ together with its recursion coefficients $\{\chi'_a, \xi'_a\}$ and $\zeta$. Hence, for a chosen spectral polynomial $q_a(z)$ with pure continuous spectrum, the matrix $\Pi$ is obtained as $\Pi = \zeta^2 I + K' - K$, where $I$ is the unit matrix and $K'$ is the same matrix as $K$ but with $\{\chi_a,\xi_a\}\mapsto\{\chi'_a,\xi'_a\}$. Consequently, the interaction vertex elements $\{\eta_{n,a,b}\}$ must be chosen to satisfy

$$\frac{1}{2}\left(\eta_{n,a,b}+\eta_{a,n,b}\right)=\sum_{c=0}^{\infty}\Pi_{b,c}(\boldsymbol{e}^{-1})_{n,a,c}=\frac{1}{2}\sum_{c=0}^{\infty}\Pi_{b,c}\left[(\boldsymbol{e}^{-1})_{n,a,c}+(\boldsymbol{e}^{-1})_{a,n,c}\right], \tag{16}$$

where the inverse tensor $\boldsymbol{e}^{-1}$ is defined by the orthogonality $\sum_{c=0}^{\infty}\boldsymbol{e}_{n,a,c}(\boldsymbol{e}^{-1})_{m,b,c}=\frac{1}{2!}\left(\delta_{n,m}\delta_{a,b}+\delta_{n,b}\delta_{m,a}\right)$. Substituting the values of the matrix elements of $\Pi$ in (16), we obtain

$$\eta_{n,a,b}=\left(\zeta^2+\tilde{\chi}_b\right)(\boldsymbol{e}^{-1})_{n,a,b}+\tilde{\xi}_{b-1}(\boldsymbol{e}^{-1})_{n,a,b-1}+\tilde{\xi}_b(\boldsymbol{e}^{-1})_{n,a,b+1}\,. \tag{17}$$

Consistency requires that in spite of the presence of the nonlinear interaction $\mathscr{L}_I$, the scalar particle corresponding to the quantum field $\Phi$ (the universal particle, which is analogous to the Higgs particle) remains massless. Now, the change in the recursion relation of the polynomial $p_n(z)$ produced by the wave equation for the quantum field $\Phi$, which contains the two terms $\eta\otimes(\bar{\Theta}\Theta)+\kappa\otimes(\bar{\Phi}\Phi)\bar{\Phi}$ added to the massless free Klein-Gordon equation, reads as follows in the space of spectral polynomials

$$\Delta p_n(z)=\sum_{a,b=0}^{\infty}\eta_{n,a,b}q_a(z)q_b(z)+\sum_{m,k,l=0}^{\infty}\kappa_{n,m,k,l}\,p_m(z)p_k(z)p_l(z)\,. \tag{18}$$

For the universal particle associated with $\Phi$ to remain massless, we require that the modified spectral polynomials $p'_n(z)$ has a pure continuous spectrum and satisfies the recursion relation (4) with $z=E^2/\lambda^2$ and $\{\alpha_n,\beta_n\}\mapsto\{\alpha'_n,\beta'_n\}$. Using Theorem 1 in Sec. 2 of Ref. [20], we can write

$$p_m(z)p_k(z)p_l(z)=\sum_{j=0}^{m+k+l}\boldsymbol{c}_{m,k,l,j}\,p_j(z)\,, \tag{19}$$

where $\boldsymbol{c}_{m,k,l,j}=\left[p_m(J)p_k(J)\right]_{l,j}$ with $J$ being the symmetric tridiagonal matrix whose elements are $J_{n,m}=\alpha_n\delta_{n,m}+\beta_{n-1}\delta_{n-1,m}+\beta_n\delta_{n+1,m}$ and $\{\alpha_n,\beta_n\}$ are the recursion coefficients for the spectral polynomials $\{p_n(z)\}$. Note that the tensor $\boldsymbol{c}$ is symmetric under the exchange of any two indices. Moreover, using Theorem 2 in Sec. 3 of Ref. [20], we can write

$$q_a(z)q_b(z)=\sum_{j=0}^{a+b}\boldsymbol{d}_{a,b,j}\,p_j(z)\,, \tag{20}$$

where $\boldsymbol{d}_{a,b,j}=\left[q_a(J)q_b(J)\right]_{0,j}=\boldsymbol{d}_{b,a,j}$ and Eq. (18) becomes $\Delta p_n(z)=\sum_{j=0}^{\infty}\Lambda_{n,j}\,p_j(z)$ with

$$\Lambda_{n,j}=\sum_{a,b=0}^{\infty}\eta_{n,a,b}\boldsymbol{d}_{a,b,j}+\sum_{m,k,l=0}^{\infty}\kappa_{n,m,k,l}\boldsymbol{c}_{m,k,l,j}\,. \tag{21}$$

Now, we require that $\Lambda$ becomes a tridiagonal symmetric matrix *without* an offset on the diagonal. That is, $\Lambda_{n,m}=\tilde{\alpha}_n\delta_{n,m}+\tilde{\beta}_{n-1}\delta_{n-1,m}+\tilde{\beta}_n\delta_{n+1,m}$. Therefore, Eq. (11) and Eq. (18) show that the modified spectral polynomial $\{p'_n(z)\}$ satisfies a three-term recursion relation whose spectral parameter is $z = E^2/\lambda^2$ and with recursion coefficients $\{\alpha'_n, \beta'_n\}$ where $\alpha'_n = \alpha_n + \tilde{\alpha}_n$ and $\beta'_n = \beta_n + \tilde{\beta}_n$. Moreover, we require that the spectrum of the polynomial $p'_n(z)$ be purely continuous. In practice, the massless universal scalar particle, which is analogous to the Higgs, is pre-determined giving the associated spectral polynomial $p'_n(z)$ along with its own recursion coefficients $\{\alpha'_n, \beta'_n\}$. Therefore, we can calculate the matrix $\Lambda$ for a given spectral polynomial $p_n(z)$ whose recursion coefficients are $\{\alpha_n, \beta_n\}$ as $\Lambda = J' - J$, where $J'$ is the same matrix as $J$ but with $\{\alpha_n, \beta_n\} \mapsto \{\alpha'_n, \beta'_n\}$. Consequently, Eq. (21) gives the required coupling elements of the vertex tensor $\{\kappa_{n,m,k,l}\}$ using this matrix $\Lambda$ and the vertex elements $\{\eta_{n,a,b}\}$ given by Eq. (17) as follows

$$\kappa_{n,m,k,l} = \sum_{j=0}^{\infty} \Lambda_{n,j} (\boldsymbol{c}^{-1})_{m,k,l,j} - \sum_{a,b,j=0}^{\infty} \eta_{n,a,b} \boldsymbol{d}_{a,b,j} (\boldsymbol{c}^{-1})_{m,k,l,j} \,. \tag{22}$$

where the inverse tensor $\boldsymbol{c}^{-1}$ is defined by the orthogonality $\sum_{l=0}^{\infty} \boldsymbol{c}_{n,m,k,l} (\boldsymbol{c}^{-1})_{n',m',k',l} = \frac{1}{3!} \delta_{(n,n'} \delta_{m,m'} \delta_{k,k')}$ with the parenthesized indices denoting permutations of the indices in accordance with the symmetry of $\boldsymbol{c}$. Substituting the values of the matrix elements of $\Lambda$ in (22), we obtained

$$\kappa_{n,m,k,l} = \tilde{\alpha}_n (\boldsymbol{c}^{-1})_{m,k,l,n} + \tilde{\beta}_{n-1} (\boldsymbol{c}^{-1})_{m,k,l,n-1} + \tilde{\beta}_n (\boldsymbol{c}^{-1})_{m,k,l,n+1} - \sum_{a,b,j=0}^{\infty} \eta_{n,a,b} \boldsymbol{d}_{a,b,j} (\boldsymbol{c}^{-1})_{m,k,l,j} \,. \tag{23}$$

Therefore, with the coupling tensors $\{\eta, \kappa\}$ given by Eq. (17) and Eq. (23), the interaction Lagrangian model $\mathscr{L}_I = \eta \otimes \Phi(\bar{\Theta}\Theta) + \frac{1}{2} \kappa \otimes (\bar{\Phi}\Phi)^2$ for the massless scalar $\Phi$ and massless spinor $\Theta$ will lead to mass generation for the spinor while keeping the scalar massless. We plan to follow this introductory theoretical work by an illustrative example showing details of the scheme for a given physical system. Nonetheless, in the Appendix we give the elements needed for the construction of such a system. We conclude the work in the following section and set forth a conjecture regarding dark energy and dark matter.

# 3. Conclusion and a conjecture

In this work, we addressed the following question: Can a massless particle acquire mass by a special nonlinear coupling to another massless particle within SAQFT? We showed that the answer is affirmative if certain conditions on the coupling parameters of the interaction vertices are satisfied, and presented a model for such an interaction. We also showed that one of the two particles (the scalar particle) remains massless and is considered the universal particle that generates mass for other elementary particles in analogy with the Higgs particle in the standard model. For simplicity, we can make the model less sumptuous by restricting to diagonal coupling in which the elements of the coupling tensors read $\eta_{n,a,b} = \eta_{n,a} \delta_{a,b}$ and $\kappa_{i,j,k,l} = \kappa_{i,j} \delta_{i,k} \delta_{j,l}$ resulting in the following equations that replace (17) and (23), respectively

$$\eta_{n,a} = \sum_{b=0}^{\infty}\left[\left(\zeta^2 + \tilde{\chi}_b\right)(\boldsymbol{e}^{-1})_{n,a,b} + \tilde{\xi}_{b-1}(\boldsymbol{e}^{-1})_{n,a,b-1} + \tilde{\xi}_b(\boldsymbol{e}^{-1})_{n,a,b+1}\right] \tag{24}$$

$$\kappa_{n,m} = \sum_{k=0}^{\infty}\left[\tilde{\alpha}_k(\hat{\boldsymbol{c}}^{-1})_{m,n,k} + \tilde{\beta}_{k-1}(\hat{\boldsymbol{c}}^{-1})_{m,n,k-1} + \tilde{\beta}_k(\hat{\boldsymbol{c}}^{-1})_{m,n,k+1}\right] - \sum_{a,j,k=0}^{\infty}\eta_{n,a}\hat{\boldsymbol{d}}_{a,j}(\hat{\boldsymbol{c}}^{-1})_{m,k,j}\,, \tag{25}$$

where $\hat{\boldsymbol{c}}_{m,n,k} := \left[p_m(J)p_m(J)\right]_{n,k}$ and $\hat{\boldsymbol{d}}_{a,j} := \left[q_a(J)q_a(J)\right]_{0,j}$.

The universal particle exploited in this work is of spin zero. Nonetheless, using a simple conformal degree counting for massless fields, we believe that a similar scheme could also be developed for a universal particle of spin 1 but not spin $\frac{1}{2}$. Now, the conformal degree for a massless (scalar, spinor, vector) particle is respectively $(-1,\ -3/2,\ -1)$. Consequently, with the conformal degree of the Lagrangian density in 3+1 Minkowski space-time being $-4$, we give in Table 1 a list of such possible models.

**Table 1**: Possible nonlinear coupling models of the universal massless field to matter fields. Repeated indices are summed over, $\left\{\gamma^{\mu}\right\}_{\mu=0}^{3}$ are the four Dirac gamma matrices, and the symmetric tensor $\sigma^{\mu\nu\lambda\rho} = \sigma^{\nu\mu\lambda\rho} = \sigma^{\mu\nu\rho\lambda}$. The term $(\bar{\Phi}\Phi)(\bar{\Psi}\Psi)$ in the first row could be generalized by adding $(\bar{\Phi}\Phi)(\bar{\Phi}\Psi)$ and $(\bar{\Psi}\Phi)(\bar{\Psi}\Psi)$.

| Universal field | | Matter Field | | Interaction Lagrangian ( $\mathscr{L}_I$ ) |
|---|---|---|---|---|
| Spin | Symbol | Spin | Symbol | |
| 0 | $\Phi$ | 0 | $\Psi$ | $\eta\otimes(\bar{\Phi}\Phi)(\bar{\Psi}\Psi) + \frac{1}{2}\kappa\otimes(\bar{\Phi}\Phi)^2$ |
| | | 1/2 | $\Theta$ | $\eta\otimes\Phi(\bar{\Theta}\Theta) + \frac{1}{2}\kappa\otimes(\bar{\Phi}\Phi)^2$ |
| | | 1 | $\mathcal{B}_{\mu}$ | $\eta\otimes(\bar{\Phi}\Phi)(\bar{\mathcal{B}}_{\mu}\mathcal{B}^{\mu}) + \frac{1}{2}\kappa\otimes(\bar{\Phi}\Phi)^2$ |
| 1 | $\mathcal{A}_{\mu}$ | 0 | $\Psi$ | $\eta\otimes(\bar{\mathcal{A}}_{\mu}\mathcal{A}^{\mu})(\bar{\Psi}\Psi) + \frac{1}{2}\kappa\otimes(\bar{\mathcal{A}}_{\mu}\mathcal{A}^{\mu})^2$ |
| | | 1/2 | $\Theta$ | $\eta\otimes\mathcal{A}_{\mu}(\bar{\Theta}\gamma^{\mu}\Theta) + \frac{1}{2}\kappa\otimes(\bar{\mathcal{A}}_{\mu}\mathcal{A}^{\mu})^2$ |
| | | 1 | $\mathcal{B}_{\mu}$ | $\sigma^{\mu\nu\lambda\rho}\eta\otimes(\bar{\mathcal{A}}_{\mu}\mathcal{A}_{\nu})(\bar{\mathcal{B}}_{\lambda}\mathcal{B}_{\rho}) + \frac{1}{2}\kappa\otimes(\bar{\mathcal{A}}_{\mu}\mathcal{A}^{\mu})^2$ |

Now, all observed massless elementary particles in nature are either of spin 1 (e.g., the photon and the gluon) or of spin 2 (the graviton) together with the possible inclusion of spin $\frac{1}{2}$ neutrino if it turns out to be exactly massless. No elementary massless particle of spin 0 (i.e., massless scalar) has ever been detected. Henceforth, if our proposed scheme becomes successful then the universal massless scalar field $\Phi$ introduced in this work, which generates mass for other particles, could contribute to the resolution of the outstanding problem of the missing dark energy. On the other hand, if it turns out that $\Phi$ has a very small mass, then it could account for some of the missing dark matter. It would be very interesting to devise a process within SAQFT that gives a unique signature for the universal scalar particle and subsequently design an experimental setup to look for and detect such signature.

## Appendix: An illustrative system

In this Appendix, we present an example of the physical model described in this paper by giving all the elements needed for its construction. We can choose the spectral polynomials associated with the massless universal scalar particle $\Phi$ as follows

$$p_n(z)=L_n^{\nu}(z)=\sqrt{\frac{(\nu+1)_n}{n!}}\,{}_1F_1\left(\begin{matrix}-n\\ \nu+1\end{matrix}\middle|z\right), \tag{A1}$$

$$p'_n(z)=L_n^{\nu'}(z)=\sqrt{\frac{(\nu'+1)_n}{n!}}\,{}_1F_1\left(\begin{matrix}-n\\ \nu'+1\end{matrix}\middle|z\right), \tag{A2}$$

where $L_n^{\nu}(z)$ is the normalized version of the Laguerre polynomial, $z=E^2/\lambda^2$, $(\nu,\nu')>-1$ and $(x)_n:=\Gamma(n+x)/\Gamma(x)$. Moreover, the associated recursion coefficients are

$$\alpha_n=2n+\nu+1,\ \beta_n=-\sqrt{(n+1)(n+\nu+1)}\,, \tag{A3}$$

$$\alpha'_n=2n+\nu'+1,\ \beta'_n=-\sqrt{(n+1)(n+\nu'+1)}\,. \tag{A4}$$

Therefore, the tridiagonal symmetric matrix $J$ becomes

$$J=\begin{pmatrix}\alpha_0 & \beta_0 & & & & & \\ \beta_0 & \alpha_1 & \beta_1 & & & & \\ & \beta_1 & \alpha_2 & \beta_2 & & & \\ & & \beta_2 & \alpha_3 & \beta_3 & & \\ & & & \times & \times & \times & \\ & & & & \times & \times & \times \\ & & & & & \times & \times\end{pmatrix} \tag{A5}$$

Moreover, the matrix $\Lambda$ is obtained using $\Lambda=J'-J$, where the matrix $J'$ is the same as $J$ above except that $\{\alpha_n,\beta_n\}\mapsto\{\alpha'_n,\beta'_n\}$.

We can also choose the spectral polynomials associated with the spinor particle $\Theta$ as follows:

$$q_a(z)=S_a^{\mu}(z;\sigma,\tau)=\sqrt{\frac{(\mu+\sigma)_a(\mu+\tau)_a}{a!(\sigma+\tau)_a}}\,{}_3F_2\left(\begin{matrix}-a,\mu+i\sqrt{z},\mu-i\sqrt{z}\\ \mu+\sigma,\mu+\tau\end{matrix}\middle|1\right), \tag{A6}$$

$$q'_a(z)=S_a^{\mu'}(y;\sigma',\tau')=\sqrt{\frac{(\mu'+\sigma')_a(\mu'+\tau')_a}{a!(\sigma'+\tau')_a}}\,{}_3F_2\left(\begin{matrix}-a,\mu'+i\sqrt{y},\mu'-i\sqrt{y}\\ \mu'+\sigma',\mu'+\tau'\end{matrix}\middle|1\right), \tag{A7}$$

where $S_a^{\mu}(x;\sigma,\tau)$ is the normalized version of the continuous dual Hahn polynomial [21] (see also Appendix A in [6,7]) and ${}_3F_2\left(\begin{matrix}a,b,c\\ d,e\end{matrix}\middle|x\right):=\sum_{n=0}^{\infty}\frac{(a)_n(b)_n(c)_n}{(d)_n(e)_n}\frac{x^n}{n!}$ is the generalized hyper-

geometric series. Moreover, $z = E^2/\lambda^2$, $y = (E^2 - M^2)/\lambda^2$, $M = \lambda\zeta$, and the associated recursion coefficients are

$$\chi_a = (a+\mu+\sigma)(a+\mu+\tau) + a(a+\sigma+\tau-1) - \mu^2, \tag{A8}$$

$$\xi_a = -\sqrt{(a+1)(a+\sigma+\tau)(a+\mu+\sigma)(a+\mu+\tau)}\,. \tag{A9}$$

where the parameters $\{\mu,\sigma,\tau\}$ are such that $\mu > 0$ and $\mathrm{Re}(\sigma,\tau) > 0$ with non-real parameters occurring in conjugate pairs. On the other hand, the parameters $\{\mu',\sigma',\tau'\}$ can assume one of two scenarios:

(1) $\mu' > 0$ and $\mathrm{Re}(\sigma',\tau') > 0$ with non-real parameters occurring in conjugate pairs, which corresponds to a structureless massive spinor $\Theta$, or

(2) $\mu' < 0$ and $\mu' + \sigma'$, $\mu' + \tau'$ are positive or a pair of complex conjugates with positive real parts. This corresponds to a massive spinor $\Theta$ with a structure of size $N+1$, where $N$ is the largest integer less than $-\mu'$.

Now, the matrix $\Pi$ is obtained as $\Pi = \zeta^2 I + K' - K$, where the tridiagonal symmetric matrices $K$ and $K'$ are the same as $J$ and $J'$ above except that $\{\alpha_n,\beta_n\} \mapsto \{\chi_a,\xi_a\}$.

Using the matrix $\Pi$ and knowledge of the spectral polynomial $p_n(z)$ given above, which enters in the calculation of $\boldsymbol{e}$, the vertex elements $\eta_{n,a,b}$ are obtained by Eq. (17) with $\boldsymbol{e}_{n,a,c} = \left[L_n^\nu(K)\right]_{a,c}$. Next, we calculate the vertex elements $\kappa_{n,m,k,l}$ via Eq. (23) using the matrix $\Lambda$ and $\eta_{n,a,b}$ just obtained and with $\boldsymbol{c}_{m,k,l,j} = \left[L_m^\nu(J)L_k^\nu(J)\right]_{l,j}$, $\boldsymbol{d}_{a,b,m} = \left[S_a^\mu(J;\sigma,\tau)S_b^\mu(J;\sigma,\tau)\right]_{0,m}$.

The physical parameters of this system that are associated with the massive spinor particle $\Theta$ are $\{M,\mu',\sigma',\tau'\}$ whereas the physical parameter $\nu'$ is associated with the universal massless scalar particle $\Phi$. On the other hand, the model parameters are $\{\lambda,\mu,\sigma,\tau,\nu\}$ with $\zeta = M/\lambda$. However, for a structureless massive spinor, the number of these parameters could be reduced significantly. For example, we could take $p_n(z) = L_n^{\pm1/2}(z)$, $p_n'(z) = L_n^{\mp1/2}(z)$, $q_a(z) = L_a^\nu(z)$, and $q_a'(z) = L_a^{\nu'}(y)$. For this alternative model, the number of physical parameters is reduced from 5 to 2 as $\{M,\mu',\sigma',\tau',\nu'\} \mapsto \{M,\nu'\}$ (e.g., mass and spin) whereas the number of model parameters are reduced from 5 to 2 as $\{\lambda,\mu,\sigma,\tau,\nu\} \mapsto \{\lambda,\nu\}$. Note that $L_n^{-1/2}(x^2)$ and $x\,L_n^{+1/2}(x^2)$ are proportional to the Hermite polynomials $H_{2n}(x)$ and $H_{2n+1}(x)$, respectively. This implies the existence of two types of universal massless scalar particles in this alternative model corresponding to $p_n'(z) = L_n^{\mp1/2}(z)$, which we may refer to as the *even* and *odd* universal particle.